\documentclass[a4paper, amsfonts, amssymb, amsmath, reprint, showkeys, nofootinbib, twoside]{revtex4-1}
\usepackage[english]{babel}
\usepackage[utf8]{inputenc}
\usepackage[colorinlistoftodos, color=green!40, prependcaption]{todonotes}
\usepackage{amsthm}
\usepackage{mathtools}
\usepackage{physics}
\usepackage{xcolor}
\usepackage{graphicx}
\usepackage[left=23mm,right=13mm,top=35mm,columnsep=15pt]{geometry} 
\usepackage{adjustbox}
\usepackage{placeins}
\usepackage[T1]{fontenc}
\usepackage{lipsum}
\usepackage{csquotes}
\usepackage{subfig}
\usepackage[pdftex, pdftitle={Article}, pdfauthor={Author}]{hyperref} 
\begin{document}
\title{Experimental observation of real spectra in Parity-Time symmetric ZRC dimers with positive and negative frequencies }

\author{St\'{e}phane Boris  Tabeu ${}^{1,2,3}$}
\email[Correspondence email address: ]{stephaneboris@yahoo.fr}
\author{ Fernande Fotsa-Ngaffo${}^{4}$}
\author{ Senghor Tagouegni ${}^{1,2}$}
\author{Kazuhiro Shouno ${}^{5}$}
\author{ Aur\'{e}lien  Kenfack-Jiotsa${}^{2,6}$}

    \affiliation {1\ Laboratory of Mechanics, Materials and Structures, Department of Physics, Faculty of Science,University of Yaounde I, P.O. Box 812,Yaounde, Cameroon } 
\affiliation {2\ Nonlinear Physics and Complex Systems Group, Department of Physics, The Higher Teacher's Training College, University of  Yaounde I, P.O. Box  47, Yaounde, Cameroon }
\affiliation {3\ Laboratory of Electronics and Signal Processing, Department of Electrical and Telecommunication Engineering, National Advanced School of Engineering,University of Yaounde I, P.O. Box  8390, Yaounde,Cameroon }
\affiliation{4\ Institute of Wood technologies, University of Yaounde I, P.O. Box 306, Mbalmayo, Cameroon}
 \affiliation{5\ Graduate School of Systems and Information Engineering, University of Tsukuba, Ibaraki 305-8573, Japan}
 \affiliation {6\ High-Tech and Slow Technology (HITASTEC), P.O. Box 105, Yaounde,
Cameroon}

\date{\today} 

\begin{abstract}
We present in this work the first experimental observation of
oscillations in Parity-Time symmetric ZRC dimers. The system
obtained is of first order ordinary differential equation  due to
the use of imaginary resistors. The coupled cells must share the
same type of frequency: positive or negative. We observed the real
and imaginary parts of the voltage across the components of a ZRC
cell. Exceptional points are well identified. This work may be
very useful in the generation of new type of oscillators. It can
also be used in the design of new optoelectronic devices for major
applications in the transport of information and the mimics of
two-level systems for quantum computing.

\end{abstract}

\keywords{Non-Hermitian systems, Parity-time symmetry, Imaginary resistor, ZRC-Dimer, Negative frequency }

\maketitle

\section{\label{intro} Introduction}
Non-Hermitian systems are today a major fields of investigations
in different specialities. In 1998, Carl Bender demonstrated that
a non-Hermitian systems can have a real spectra without being
Hermitian opening the road to Parity-Time (PT) Symmetry
\cite{a1,a2,a3}. In Quantum Mechanics,  this reality is related to
the potential which satisfied $V\left( x \right) = {V^*}\left( { -
x} \right)$ with an even real part and odd imaginary part
\cite{a1,a2,a3,a4,a5,a6}. In Optics, the refractive index of the
medium must satisfied a relation of the same nature $n\left( x
\right) = {n^*}\left( { - x} \right)$
\cite{a7,a8,a9,a10,a11,a12,a13} and $\varepsilon \left( x \right)
= {\varepsilon ^*}\left( { - x} \right)$ for dielectric
permittivity in Metamaterials \cite{a14,a15,a16,a17}. The notion
of Parity-Time Symmetry is also extended to many order fields such
as Photonics \cite{a7,a8,a9,a10,a11,a12,a13}, Mechanics
\cite{a18,a19,a20}, Optomechanics \cite{a20,a21,a22,a23},
Acoustics \cite{a24,a25,a26} and Electronics
\cite{a27,a28,a29,a30,a31} just to name few. In 2010, Reuter
realized the first experimental demonstration of PT-symmetry in
Optics by coupling two waveguides having equal amount of gain and
loss \cite{a7}. In 2011, Schindler et al used two active RLC
circuits to introduce the notion in Electronics \cite{a27,a28}.
The loss cell was represented by a natural positive resistor and
the gain cell by a negative resistor using Negative impedance
converter. In 2013, the team of Carl Bender experimentally
demonstrate the real spectra in coupled mechanic oscillators
\cite{a18}. The Rabi oscillations occurred near the the
exceptional point in the response of the system. More after it was
presented the counterpart of Parity-Time symmetry: The Anti-Parity
Time symmetry. In this new Physics, the loss and the gain are
exclusively in the dimers \cite{a30,a32,a33,a34,a35,a36,a37,a38}.
In Quantum Mechanics, the potential follows  the relation $V(x) =
- {V^*}\left( -x \right)$ , the refractive index  $n(x) = -
{n^*}\left( -x \right)$ in Optics and $\varepsilon \left( x
\right) =  - {\varepsilon ^*}\left( { - x} \right)$ for the
dielectric permittivity in Metamaterials. Now the real parts of
the potential, the refractive index of the medium and the
permittivity are odd and their imaginary parts even giving rise to
the coupled of Gain- Gain cells or Loss-Loss cells to made
Anti-Parity Time symmetry. In others fields, the rotating frame is
used to create indirectly  positive and negative frequencies in
the cells of the dimer such as in lasers gyroscopes \cite{a36}. It
also used the Coupled Mode Theory (CMT) and adiabatic elimination
in optical waveguides , slowly varying envelope approximation in
Nanophotonics  to achieve the same reality. Many others works are
also presented in these directions
\cite{a30,a31,a32,a33,a34,a35,a36,a37}. Tabeu et al presented in
\cite{a30} how to achieve Parity-Time symmetry and Anti-Parity
time symmetry in electronics using imaginary resistors
\cite{a31,a39,r1} to have directly the system of first order
ordinary  differential equation without the coupling mode theory
or the rotating frame. The configurations presented were with
Gain-Loss, Gain-Gain and Loss-Loss in the dimer. All these
(Anti)-Parity-time systems have many applications such as sensing
and telemetry, non-reciprocal transport of information, generation
of qubits for quantum computing, wireless power transfer,
switches,  single mode power Lasers
\cite{a9,a10,a11,a12,a17,a30,a31,a34,a38,r1,a40,a41,a42,a43,a44,a45}
and others .

In this works, we proposed the experimental realization of ZRC
cells and a PT-dimer based imaginary resistors. Since the
Hamiltonian of the resulted system is the same which the one
encounter after some transformations in Optics and Photonics, it
can mimics easily such these systems and gather the bridge for new
devices in Optolectronics and Photoelectronics with the avenue of
optical and photonic quantum computing. The paper is organized as
follows: In Sect.~\ref{sec:2} the RLC and ZRC are presented with
their frequencies. In Sect.~\ref{sec:3}, the ZRC PT-dimer is
studied. Its eigenvalues are calculated and some simulations of
its dynamic behavior are driven. In Sect.~\ref{sec:4}, the
experimental realization of the dimer is presented and the
oscillations across the gain and loss cells. The paper end with a
conclusion.

\section{RLC and ZRC Cells}
\label{sec:2}
\subsection{Real volatge in RLC circuit}

\begin{figure}[!ht]
\centering
\subfloat(a){\includegraphics[width=1.25in,height=1.25in]{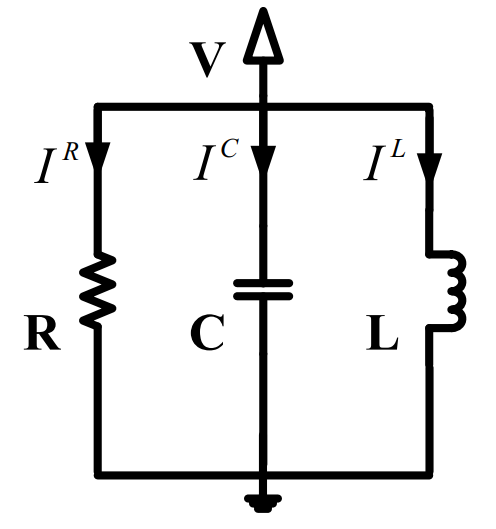}}
\subfloat(b){\includegraphics[width=1.25in,height=1.25in]{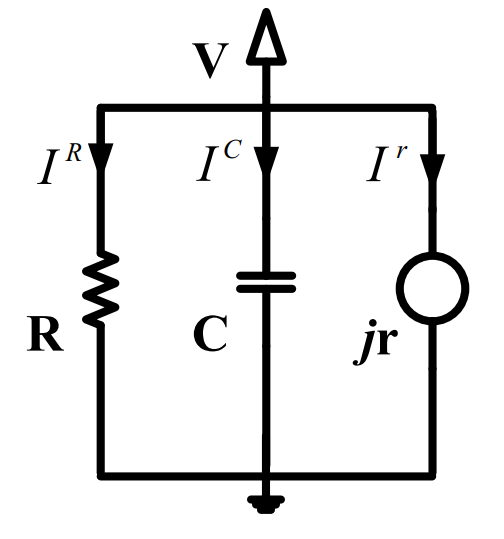}}
\caption{(a) The RLC cell is constituted of a real resistor $R$,
an inductor $L$ and a capacitor $C$. All the components are
mounted in parallel. The ZRC cell is constituted of an imaginary
resistor $Z=jr$, a real resistor $R$ and a capacitor $C$ all
mounted in parallel.  }
\label{fig:1}
\end{figure}

The RLC cell contains a real resistor,  a capacitor and an
inductor Fig.\ref{fig:1}(a) The Kirchhoff's
 law applied at the node
gives us:
\begin{equation}
{I^R} + {I^L} + {I^C} = 0
\label{eq:1}
\end{equation}
By applying the Ohm's law at the borders of the components, we
have a second order form of ordinary differential equation
depicted as :
\begin{equation}
\frac{{{d^2}V}}{{d{t^2}}} + \gamma \frac{{dV}}{{dt}} + \omega
_0^2V = 0{\rm{ }} \label{eq:2}
\end{equation}
With $\gamma  = \frac{1}{{RC}}$ representing the damping rate and
${\omega _0} = \sqrt {\frac{1}{{LC}}} $  the natural frequency of
the cell. We look a solution in the form  $V \propto {e^{j\alpha
t}}$  and it derives from Eq. (~\ref{eq:2})  a second order
equation in $ \alpha $ :
\begin{equation}
{\alpha ^2} - j\gamma   - \omega _0^2 = 0 \label{eq:3}
\end{equation}
The final voltage with an appropriate choice of initial conditions
is :
\begin{equation}
V(t) = \frac{1}{2}{V_0}{e^{ - \left( {\frac{\gamma }{2}}
\right)t}}\left( {{e^{j\omega t}} + {e^{ - j\omega t}}} \right)
\label{eq:4}
\end{equation}

\begin{equation}
{\alpha _{1,2}} = j\left( {\frac{\gamma }{2}} \right) \pm \sqrt
{\omega _0^2 - {{\left( {\frac{\gamma }{2}} \right)}^2}}  = 0
\label{eq:5}
\end{equation}
with $\omega  = \sqrt {\omega _0^2 - {{\left( {\frac{\gamma }{2}} \right)}^2}}$ .\\
The system is under-damped if $\omega  > 0$  . Then, the system is
pseudo-oscillatory with exactly the combination of two voltages
with positive and negative frequencies:
\begin{equation}
V(t) = {\bf{V}}\left( \omega  \right) + \,{\bf{V}}\left( { -
\omega } \right)\ \label{eq:6}
\end{equation}
The result is exactly a real value.
\begin{equation}
V(t) = {V_0}{e^{ - \left( {\frac{\gamma }{2}} \right)t}}\cos
\left( {\omega t} \right) \label{eq:7}
\end{equation}

\subsection{Complex voltage in ZRC circuit}

\begin{figure*}[!htbp]
\centering
\includegraphics[width=6.5in,height=5in]{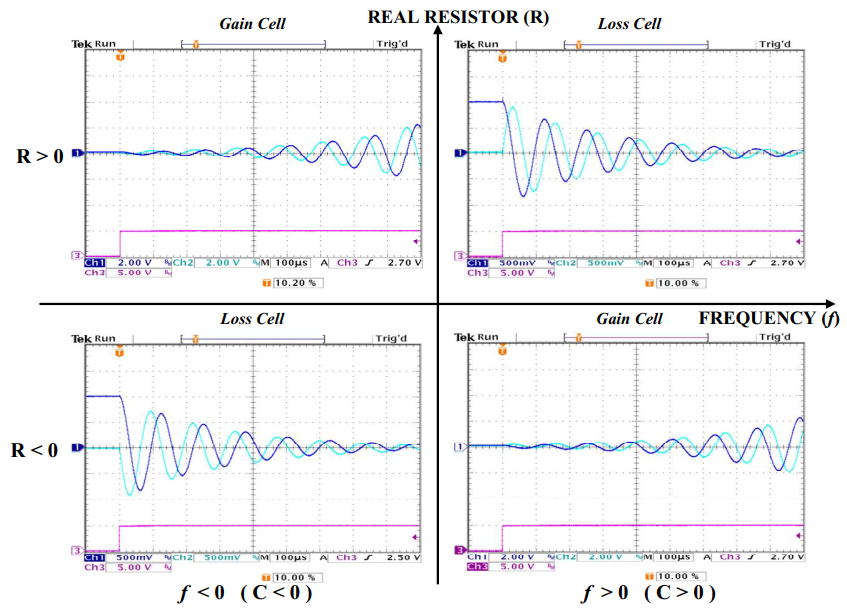}
\caption{Experimental realization of ZRC circuits with positive
and negative frequencies. Pseudo-oscillations of the real $\Re
\left( {{\bf{V}}\left( \omega  \right)} \right) = {V_0}{e^{\left(
{ - \gamma t} \right)}}\cos \left( {\omega t} \right)$ (in blue)
and the imaginary $\Im \left( {{\bf{V}}\left( \omega  \right)}
\right) = {V_0}{e^{\left( { - \gamma t} \right)}}\sin \left(
{\omega t} \right)$ (in cyan) parts of the voltage . The sign of
the frequency is the one of the capacitors. the experiments run a
$f \approx  \pm {\rm{ }}7.96{\rm{ }}kHz$ . The others values of
components are : $Z = j2{\rm{ }}k\Omega \,$ for the imaginary
resistor ;$ R =  \pm 30\,k\Omega $ for the real resistors and $C =
\pm 10{\rm{ }}nF$ for capacitors}
 \label{fig:2}
\end{figure*}

\begin{figure}[!ht]
\centering
\includegraphics[width=3.25in,height=2in]{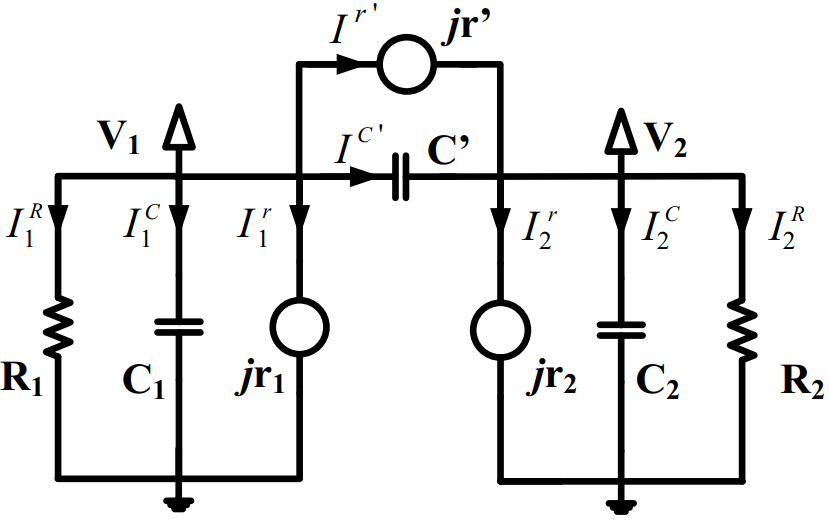}\\
\caption{The circuit of the ZRC Dimer.  Each cell contains in
parallel an imaginary resistor ${Z_n} = j{r_n}$, a real resistor
${R_n}$, a capacitor ${C_n}$. The  ZRC-cells are coupled by an
imaginary resistor $z=jr'$ or a capacitor $C'$ or by all of them.}
\label{fig:3}
\end{figure}

\begin{figure}[!ht]
\centering
\subfloat(a){\includegraphics[width=3in,height=2.5in]{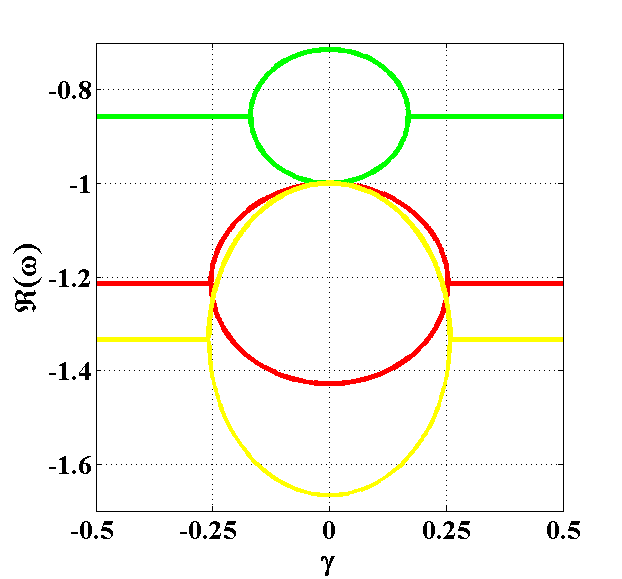}}
\subfloat(b){\includegraphics[width=3in,height=2.5in]{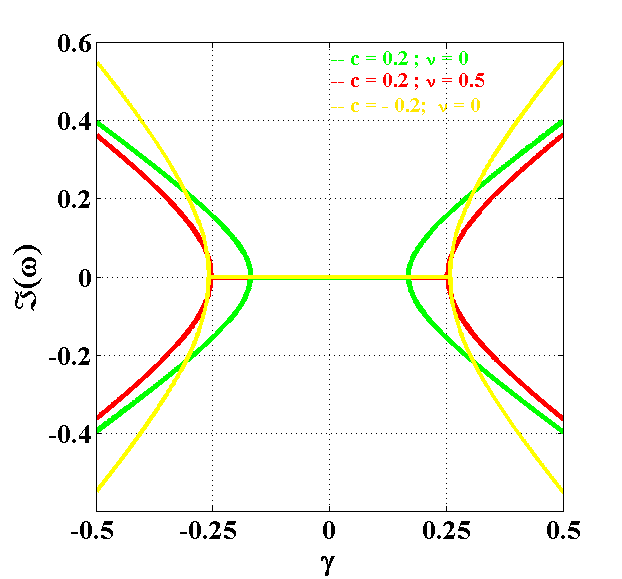}}
 \caption{ The eigenvalues of the setup corresponding to three cases in which we have a positive and
negative capacitive couplings and at last a case with both the
capacitive and the imaginary couplings. (a) Real parts of the
eigenvalues (b) Imaginary parts of the eigenvalues } \label{fig:4}
\end{figure}

\begin{figure}[!ht]
\centering
\includegraphics[width=3.5in,height=2.5in]{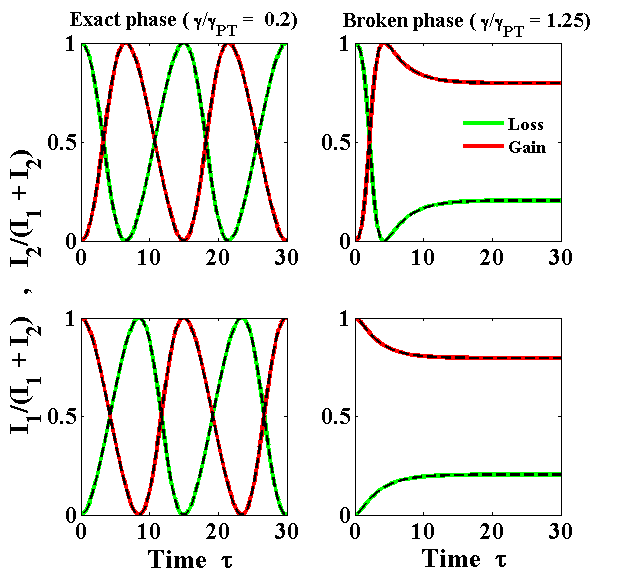}
\caption{Relative values of the modules of complex voltages across
the gain and loss cells ${I_n} = {\left| {{V_n}\left( \tau
\right)} \right|^2}$ with different initial inputs. In the exact
phase, there are normal oscillations with an amplification of the
initial input due the presence of gain and loss in the setup. In
the broken phase, there is an exponential growth and also a deep
difference between the behavior of gain and loss cells }
\label{fig:5}
\end{figure}

\begin{figure*}[!htbp]
 \centering
\subfloat(a){\includegraphics[width=3in,height=2.5in]{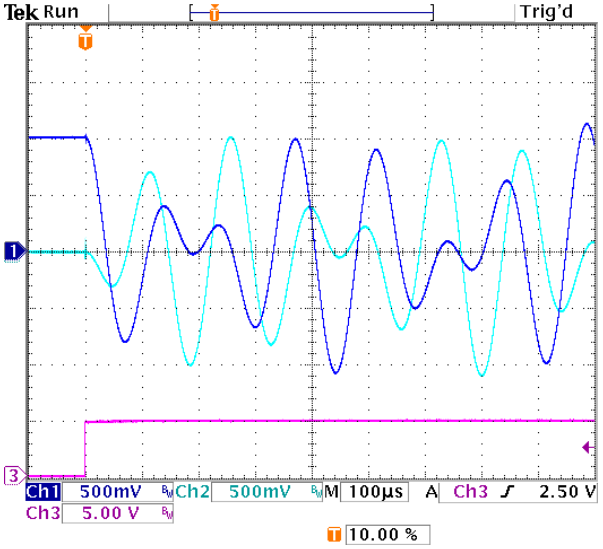}}
\subfloat(b){\includegraphics[width=3in,height=2.5in]{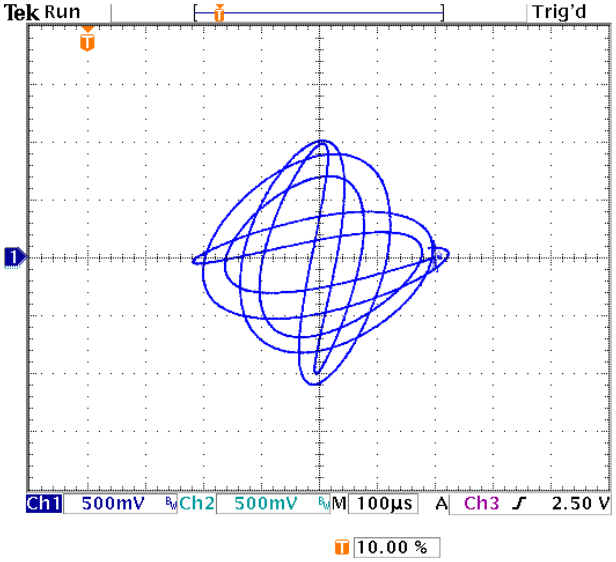}}\\
\subfloat(c){\includegraphics[width=3in,height=2.5in]{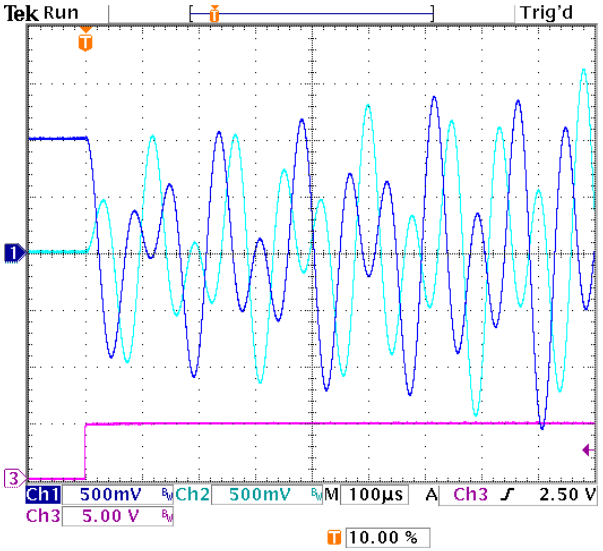}}
\subfloat(d){\includegraphics[width=3in,height=2.5in]{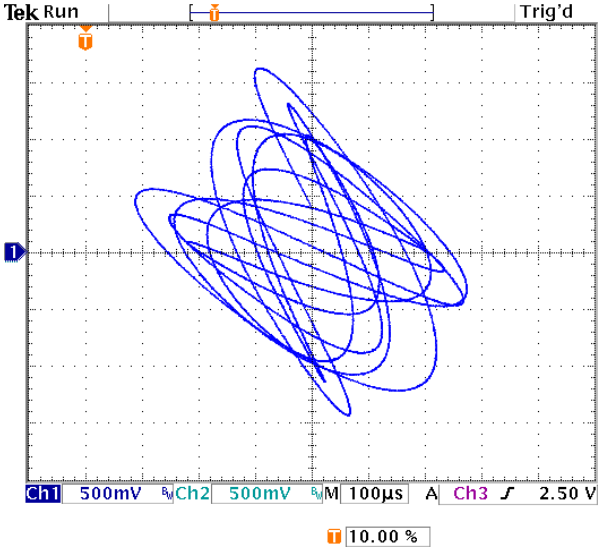}}\\
\subfloat(e){\includegraphics[width=3in,height=2.5in]{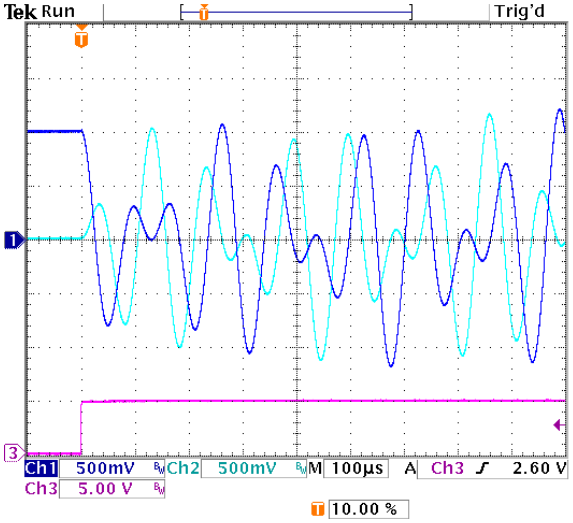}}
\subfloat(f){\includegraphics[width=3in,height=2.5in]{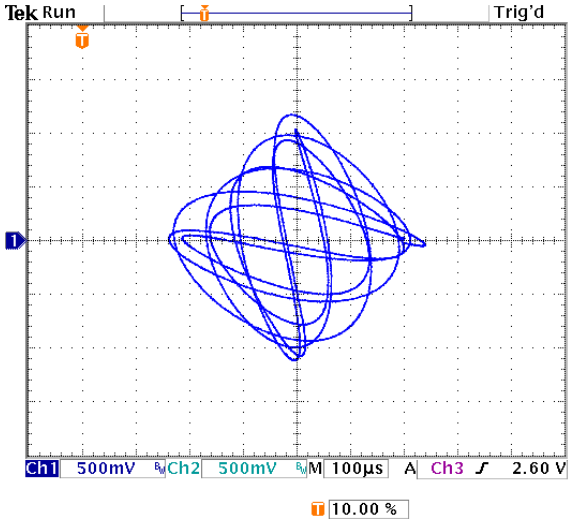}}\\
\caption{ Experimental observation  of real part of voltages in
the ZRC PT-dimer at $\gamma  = 0.2\,{\gamma _{PT}}$ . The first
case with positive frequency and capacitive coupling$ C = {\rm{
}}10{\rm{ }}nF\,;\,\,C' = \,\,2\,nF\,;{\rm{ }}Z = j2{\rm{
}}k\Omega  ;\,\,c = {\rm{ }}0.2\,\,;\,\,\nu  = {\rm{ }}0$:  (a)
The real voltage dynamic across the loss and the gain cells
$\left( {f \approx 2.23\,kHz\,\,;\,\,R \approx 59.16\,k\Omega }
\right)$ (b)  The Lissajous's curve. The second case with negative
frequency and capacitive coupling $ C = {\rm{ }}-10{\rm{
}}nF\,;\,\,C' = \,\,2\,nF\,;{\rm{ }}Z = j2{\rm{ }}k\Omega  ;\,\,c
= {\rm{ }}-0.2\,\,;\,\,\nu  = {\rm{ }}0$:  (c) The real voltage
dynamic across the loss and the gain cells $\left( {f \approx
-5.2\,kHz\,\,;\,\,R \approx 38.73\,k\Omega } \right)$ (d)  The
Lissajous's curve. The third case with positive frequency ,
capacitive and imaginary couplings $ C = {\rm{ }}10{\rm{
}}nF\,;\,\,C' = \,\,2\,nF\,;{\rm{ }}Z = j2{\rm{ }}k\Omega  ;\,\,c
= {\rm{ }}0.2\,\,;\,\,\nu  = {\rm{ }}0.5$:  (e) The real voltage
dynamic across the loss and the gain cells $\left( {f \approx
3.34\,kHz\,\,;\,\,R \approx 39.44\,k\Omega } \right)$ (f)  The
Lissajous's curve.}
 \label{fig:6}
\end{figure*}

\begin{figure}[!htbp]
\centering
\includegraphics[width=3.25in,height=2in]{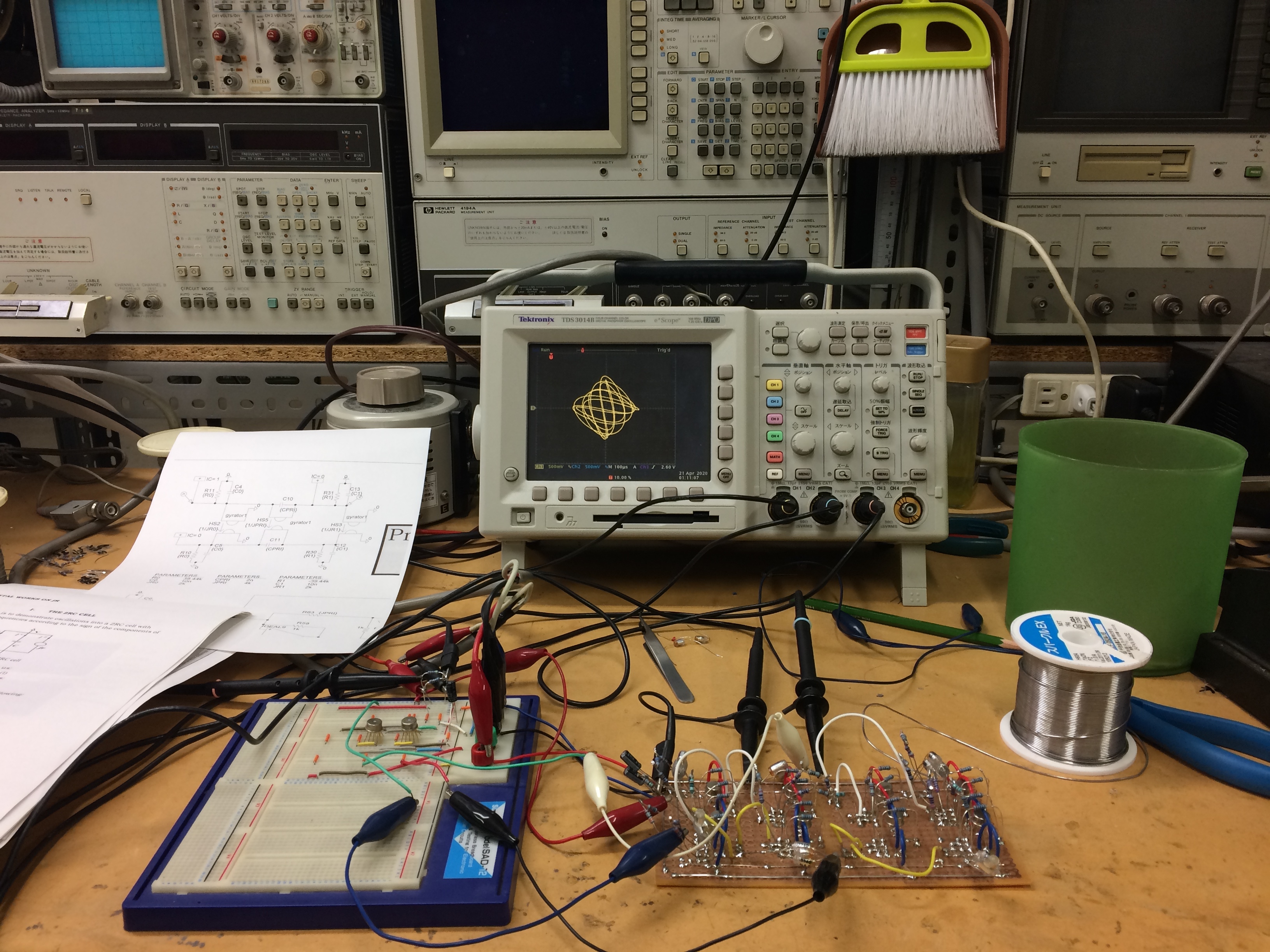} \caption{View of the experiment
environment}
 \label{fig:7}
\end{figure}


The imaginary resistor is a resistor which resistance is a pure
imaginary number. It is built indirectly by gyrators \cite{a39}.
If it is associated to a resistor which resistance is a real
number, we can have a complex resistor which impedance is in the
form ${R_C} = R + jr = \left| {{R_C}} \right|{e^{j\Phi }}$ . $ R $
represents the real part and $ r $ the imaginary part $\left(
{\left| {{R_C}} \right| = \sqrt {{R^2} + {r^2}} \,\,;\tan \left(
\Phi \right) = \frac{r}{R}} \right)$ . The ZRC cell  is
constituted by a capacitor $C$  , a real resistor $R$   and at
least an imaginary resistor $Z=jr$ \cite{a30,a31}.    The value of
components may be positive or negative without any restrictions.
The cell is depicted in Fig.~\ref{fig:1}(b) . The Kirchhoff's law
applied at the node gives us:
 \begin{equation}
{I^Z} + {I^R} + {I^C} = 0
\label{eq:8}
\end{equation}
The Ohm's law at the borders of resistors involves:
  \begin{equation}
V = R{I^R} = Z{I^Z}
\label{eq:9}
\end{equation}

From the combination of Eq.~\ref{eq:8} and Eq.~\ref{eq:9}, we
deduce the following:
 \begin{equation}
C\frac{{dV}}{{dt}} + \frac{V}{R} - j\frac{V}{r} = 0 \label{eq:10}
\end{equation}

  \begin{equation}
\frac{{dV}}{{dt}} + \left( {\frac{1}{{RC}} - j\frac{1}{{rC}}}
\right)V = 0
\label{eq:11}
\end{equation}
The final voltage with an appropriate initial conditions is:
\begin{equation}
V\left( t \right) = {V_0}{{\mathop{\rm e}\nolimits} ^{\left( { -
\frac{1}{{RC}}t} \right)}}{{\mathop{\rm e}\nolimits} ^{\left(
{j\frac{1}{{rC}}t} \right)}}
\label{eq:12}
\end{equation}
We obtained a complex voltage across the components of the circuit
given as:
\begin{equation}
V\left( t \right) = {V_0}{e^{\left( { - \gamma t}
\right)}}{e^{\left( {j\omega t} \right)}}{\rm{  =
}}{V_0}{e^{\left( { - \gamma t} \right)}}\left( {\cos \left(
{\omega t} \right) + j\sin \left( {\omega t} \right)} \right)
\label{eq:13}
\end{equation}
with $\gamma  = \frac{1}{{RC}}$  and
$\omega  = \frac{1}{{rC}}$  which may be positive or negative.\\
The system with positive frequency does not describe the same
reality with the one with negative frequency because $\cos \left(
{\omega t} \right)$  is even and $\sin \left( {\omega t} \right)$
is odd.
\begin{equation}
\left\{ \begin{array}{l}
{\bf{V}}\left( \omega  \right) = {V_0}{e^{\left( { - \gamma t} \right)}}\left( {\cos \left( {\omega t} \right) + j\sin \left( {\omega t} \right)} \right)\\
{\bf{V}}\left( { - \omega } \right) = {V_0}{e^{\left( { - \gamma
t} \right)}}\left( {\cos \left( {\omega t} \right) - j\sin \left(
{\omega t} \right)} \right)
\end{array} \right.
\label{eq:14}
\end{equation}

\begin{equation}
\left\{ \begin{array}{l}
\Re \left( {{\bf{V}}\left( \omega  \right)} \right) = \Re \left( {{\bf{V}}\left( { - \omega } \right)} \right) = {V_0}{e^{\left( { - \gamma t} \right)}}\cos \left( {\omega t} \right)\\
\Im \left( {{\bf{V}}\left( \omega  \right)} \right) =  - \Im
\left( {{\bf{V}}\left( { - \omega } \right)} \right) =
{V_0}{e^{\left( { - \gamma t} \right)}}\sin \left( {\omega t}
\right)
\end{array} \right.
\label{eq:15}
\end{equation}

Since this response of the system is complex, based on the model
of the imaginary resistor proposed in Ref.\cite{a39} , one can
probe the real part $\Re \left( {{\bf{V}}\left( \omega  \right)}
\right) = {V_0}{e^{\left( { - \gamma t} \right)}}\cos \left(
{\omega t} \right)$   and the imaginary part  $\Im \left(
{{\bf{V}}\left( \omega  \right)} \right) = {V_0}{e^{\left( { -
\gamma t} \right)}}\sin \left( {\omega t} \right)$  of the system
simultaneously. The sign of the damping rate define the nature of
the cell. Depending on the signs of the components,  When $\gamma
< 0$, we have a gain cell and when $\gamma > 0$, we have a loss
cell. The experimental verification of the behavior of the cell
with positive and negative frequencies is given in
Fig.~\ref{fig:2}. The real and the imaginary parts of the voltage
are presented simultaneously for each case. The sign of the
frequency is the one of the capacitor in the cell. This gives the
possibilities to achieve experimentally PT-symmetric systems with
cells of same frequency and Anti-Parity-Time (APT) symmetry with
cells of opposite frequencies without the rotating frame condition
as in existing  systems in others fields of Physics. These results
open new avenues for experimental non-Hermitian Quantum Mechanics
and Quaternionic Quantum Electronics.

\section{ZRC PARITY-TIME SYMMETRIC DIMER }
\label{sec:3}

\subsection{The ZRC model and equations of dynamics}

The ZRC PT- symmetric dimer is made by two active ZRC-cells. The
cells contain in parallel an imaginary resistor, a capacitor and a
real resistor. They are coupled by a capacitor or an imaginary
resistor  or by  all of them. The  values of components may be
positive or not without any restriction. In Fig.~\ref{fig:3} is
presented the setup of the ZRC PT-symmetric dimer. By applying the
Kirchoff's laws at nodes 1 and 2, we have:

\begin{equation}
\left\{ \begin{array}{l}
I_1^R + I_1^C + I_1^Z + I_1^G + {I^{C'}} + {I^{r'}} = 0{\rm{    }}\\
{\rm{ }}I_2^R + I_2^C + I_2^Z + I_2^G - {I^{C'}} - {I^{r'}} =
0{\rm{   }}
\end{array} \right.
\label{eq:16}
\end{equation}
with $I_n^C = {C_n}\frac{d}{{dt}}{V_n}$ ; ${I^{C'}} =
C'\frac{d}{{dt}}\left( {{V_0} - {V_1}} \right)$ ; $I_n^R =
\frac{{{V_n}}}{{{R_n}}}\,\,$; $I_n^Z =  -
j\frac{{{V_n}}}{{{r_n}}}\,\,$; ${I^{R'}} = \frac{1}{{R'}}\left(
{{V_1} - {V_2}} \right)$ ; ${I^{r'}} =  - j\frac{1}{{r'}}\left(
{{V_1} - {V_2}} \right)$ for all the  components.  These
considerations lead on the general form of first order ordinary
differential equations:

\begin{equation}
\left\{ \begin{array}{l}
\frac{{d{V_1}}}{{d\tau }} = \frac{1}{\Delta }\left[ {\left( {1 + {c_2}} \right)\left( {j\left( {1 + {\nu _1}} \right) - {\gamma _1}} \right) - j{c_1}\Gamma {\nu _2}} \right]{V_1}{\rm{  }}\\
\, + {\rm{  }}\frac{1}{\Delta }\left[ { - j\left( {1 + {c_2}} \right){\nu _1} + {c_1}\Gamma \left( {j\left( {1 + {\nu _2}} \right) - {\gamma _2}} \right)} \right]{V_2}\\
\\
\frac{{d{V_2}}}{{d\tau }} = \frac{1}{\Delta }\left[ { - j\left( {1 + {c_1}} \right)\Gamma {\nu _2} + {c_2}\left( {j\left( {1 + {\nu _1}} \right) - {\gamma _1}} \right)} \right]{V_1}{\rm{  }}\\
\, + {\rm{  }}\frac{1}{\Delta }\left[ {\left( {1 + {c_1}}
\right)\Gamma \left( {j\left( {1 + {\nu _2}} \right) - {\gamma
_2}} \right) - j{c_2}{\nu _1}} \right]{V_2}
\end{array} \right.
\label{eq:17}
\end{equation}
\\
where $\tau  = {\omega _1}t$; ${\omega _n} = \frac{1}{{{r_n}{C_n}}}$; $\Gamma  = \frac{{{\omega _2}}}{{{\omega _1}}}$; ${c_n} = \frac{{C'}}{{{C_n}}}$; ${\nu _n} = \frac{{{r_n}}}{{r'}}$; ${\gamma _n} = \frac{{{r_n}}}{{{R_n}}}$ and  $\Delta  = 1 + \sum\limits_{n = 1}^2 {{c_n}} $.\\
The system of  first order ordinary differential equation is   PT-
symmetric when the several conditions are satisfied:

\begin{equation}
\left\{ \begin{array}{l}
\Gamma  = 1{\rm{ }}\,;{\rm{ }}\,{c_1} = {c_2} = c{\rm{ }}\,\\
\,{\nu _1} = {\nu _2} = \nu \,\,\,;\,\,{\gamma _1} =  - {\gamma
_2} = \gamma \,
\end{array} \right.
\label{eq:18}
\end{equation}

 We can associate to our system a non-Hermitian effective Hamiltonian such that :

 \begin{equation}
j\frac{d}{{d\tau }}\left| {\Psi \left( \tau  \right)}
\right\rangle  = {H_{eff}}\left| {\Psi \left( \tau  \right)}
\right\rangle \label{eq:19}
\end{equation}

where: $j$  is the imaginary unit $\left( {{j^2} =  - 1} \right)$
; $\left| {\Psi \left( \tau  \right)} \right\rangle  = {\left(
{{V_1}\left( \tau  \right),{V_2}\left( \tau  \right)} \right)^T}$;
${H_{eff}} = {k_0}I + {k_x}{\sigma _x} + {k_y}{\sigma _y} +
{k_z}{\sigma _z}$ . $I$  is the $2 \times 2$ matrix unit and
${\sigma _x}$  , ${\sigma _y}$  and  ${\sigma _z}$ are Pauli's
matrices.

 \begin{equation}
{\sigma _x} = \left( {\begin{array}{*{20}{c}}
0&1\\
1&0
\end{array}} \right){\kern 1pt} {\kern 1pt} ;\,\,{\sigma _y} = \left( {\begin{array}{*{20}{c}}
0&{ - j}\\
j&0
\end{array}} \right){\kern 1pt} {\kern 1pt} ;\,\,{\sigma _z} = \left( {\begin{array}{*{20}{c}}
1&0\\
0&{ - 1}
\end{array}} \right){\kern 1pt}
\label{eq:20}
\end{equation}
 The complex coefficients ${k_0}$ ,${k_x}$,${k_y}$, and ${k_z}$  are   function of the independent real parameters of the system  deduced  as :
\begin{equation}
\left\{ \begin{array}{l}
{k_0} =  - \left( {1 + c + \nu } \right)/\Delta \,\,\,\,;\,\,\,\,{k_x} =  - \left( {c - \nu } \right)/\Delta \,\,\\
{k_y} =  - c\gamma /\Delta \,\,\,;\,\,\,\,{k_z} =  - j\left( {1 +
c} \right)\gamma /\Delta \,\,\,\,\,
\end{array} \right.
 \label{eq:21}
\end{equation}

 Since the system is PT-symmetric, it satisfied the following relation:

  \begin{equation}
\left[ {{\bf{PT}}\,,{H_{eff}}} \right] = 0
 \label{eq:22}
\end{equation}
 where ${\bf{P}} = {\sigma _x}$  is the Parity operator and ${\bf{T}} = K$ is the time reversal operator with $K$  being the complex conjugation operation. That is ${k_0}$, ${k_x}$ and  ${k_y}$     must be real whereas ${k_z}$ must be purely imaginary. The eigenvalues of the Hamiltonian resulted from $\det \left( {{H_{eff}} - \omega I} \right) = 0$  are given as follow:

 \begin{equation}
{\omega _ \pm } = {k_0} \pm \sqrt {k_x^2 + k_y^2 + k_z^2}
 \label{eq:23}
\end{equation}
\subsection{The eigenvalues of the  ZRC PT-dimer and oscillations }
 Function of the real parameters of the system,  the eigenvalues of the effective Hamiltonian are :
 \begin{equation}
{\omega _ \pm } = \frac{{ - (1 + c + \nu ) \pm \sqrt {\left( {1 +
2c} \right)\left( {\gamma _{PT}^2 - {\gamma ^2})} \right)} }}{{1 +
2c}} \label{eq:24}
\end{equation}
with ${\gamma _{PT}} =  \pm \frac{{\left| {c - \nu } \right|}}{{\sqrt {1 + 2c} }}$. \\
The system is PT-unbreakable if $c <  - {1 \mathord{\left/
 {\vphantom {1 2}} \right.
 \kern-\nulldelimiterspace} 2}$ and PT-breakable
$c >  - {1 \mathord{\left/
 {\vphantom {1 2}} \right.
 \kern-\nulldelimiterspace} 2}$. When $c = \nu $, one have a non-Hermitian diabolic point in the unbreakable Parity-Time symmetry and a thresholdless point in the breakable Parity-Time symmetry.

 The exceptional points are where  both  eigenvalues and eigenvecors coalesce. There  are well identified in Fig.~\ref{fig:4}. Three cases are presented.
 In the first case only the positive capacitive coupling is activated $\left( {c = {\rm{ }}0.2\,\,;\,\,\nu  = {\rm{ }}0} \right)$.
 When $\gamma  < {\gamma _{PT}}$, all the eigenfrequencies are real $\left( {\Re \left( {{\omega _ \pm }} \right) \ne 0\,;\,\Im \left( {{\omega _ \pm }} \right) = 0} \right)$
 and we are in the exact phase of Parity-Time Symmetry. When $\gamma  = {\gamma _{PT}}$, the eigenvalues coalesce $\left( {{\omega _ + } = {\omega _ - }} \right)$ :
  this is the exceptional point. Its presence in a system calls for   enhancement of sensing than the one with diabolic
  point, for chiral properties which are revealed when the point is encircled with time varying components in the setup,
  opening the road to explain the asymmetric transport of information in Optics and in Photonics.
  It is at this point that happens unidirectional invisibility in the scattering properties of waveguides with Parity-Time symmetry.
   At last, when $\gamma  > {\gamma _{PT}}$, the real parts of eigenvalues are identical $\left( {\Re \left( {{\omega _ + }} \right) = \Re \left( {{\omega _ - }} \right)} \right)$ and the imaginary parts emerge opposite in sign $\left( {\Im \left({{\omega _ + }} \right) =  - \Im \left( {{\omega _ - }} \right)}\right)$ :
   we are in the broken phase Parity-Time symmetry. In this phase the system is nonreciprocal and there is a deep difference between gain cell
   and loss cell when the initial input is set to one or another cell. The Coherent Perfect Absorber-Laser (CPA-L) occurs
   in the this phase when the scattering properties are investigated in PT-waveguides.
   We have the same description with the second case with negative capacitive coupling $\left( {c =  - \,0.2\,;\,\,\nu  = 0} \right)$
  but more than $-1/2$ to be in the breakable PT-symmetry and the third case with both positive capacitive coupling and
  imaginary coupling $\left( {c = \,0.2\,;\,\,\nu  = 0.5} \right)$.
  For the third case, analytical solutions  and numerical simulations with the algorithm of Runge-Kutta order 4 are presented in Fig.~\ref{fig:5}.
  The behaviors of oscillations when the  input is set from the loss is equal to the one of the initial input from gain after a beat of time.
  This confirms  the character of non-reciprocity which is a signature of non-Hermitian systems.

\section{Experimental realization of oscillations in ZRC PT-Dimer}
\label{sec:4}

 In this section is presented the experimental  realization of the ZRC Parity-Time symmetric dimer with positive and negative frequencies.
 The imaginary resistor used is  in the model of Ref.\cite{a39}. One can easily probe the real and imaginary parts of the voltages across the components.
  We have use the Negative Immitiance Converter to achieve negative values of components with resistors, capacitors and operational amplifiers.
  The realizations are those of the three configurations listed in the previous section at $\gamma  = 0.2\,{\gamma _{PT}}$.
  Only the real parts of voltages are presented in Fig.~\ref{fig:6}. In the first case, the positive natural frequency of independent cells is $f = 7.96\,kHz$.
  The values of different components are:${Z_0} = {Z_1} = Z = j2\,k\Omega \,;\,\,\,{R_0} = 59.16\,k\Omega \,\,{R_1} =  - 59.16\,k\Omega ;\,{C_0} = {C_1} = \,C = 10\,nF\,;\,\,C' = \,2\,nF\,;\,\,jR' = 0$.
  The frequency of oscillations obtained is ${f \approx 2.23\,kHz}$ (Fig.~\ref{fig:6}(a-b)).
 In the second  case, the negative natural frequency of independent cells is $f = - 7.96\,kHz$.
 The values of different components
 are:${Z_0} = {Z_1} = Z = j2\,k\Omega \,;\,\,\,{R_0} = 38.73\,k\Omega \,\,{R_1} =  - 38.73\,k\Omega ;\,{C_0} = {C_1} = \,C = -10\,nF\,;\,\,C' = \,2\,nF\,;\,\,jR' = 0$.
 The frequency of oscillations obtained is ${f \approx -5.2\,kHz}$ (Fig.~\ref{fig:6}(c-d)). In the third  case, the positive natural frequency of independent cells is $f = 7.96\,kHz$.
 The values of different components are:${Z_0} = {Z_1} = Z = j2\,k\Omega \,;\,\,\,{R_0} = 39.44\,k\Omega \,\,{R_1} =  - 39.44\,k\Omega ;\,{C_0} = {C_1} = \,C = 10\,nF\,;\,\,C' = \,2\,nF\,;\,\,jR' = j4\,k\Omega \,$.
 The frequency of oscillations obtained is ${f \approx 3.34\,kHz}$ (Fig.~\ref{fig:6}(e-f)).

 Operational amplifiers (LF356), metal-film resistors and polystyrene capacitors are  used in the experimental circuits.
 All the elements are tuned to be within $0.1\% $ with respect to the desired values.
 For injecting the initial conditions, we have used DG200ABA analog switches and and an
 external standard voltage source. All the waveforms are acquired by Tektronix TDS3014B digital oscilloscope.
 Agilent 33120A function generator is used to generate the trigger signal for the experimental
 circuits (See Fig.~\ref{fig:7})

\section{Conclusion}
We have reported in this work the experimental realization of the
ZRC PT-symmetric Dimer. A comparative study is made between RLC
and ZRC cells. Contrary to the RLC cell which dynamic is of second
order ordinary differential equation with a real output, the ZRC
is of first order ordinary differential equation with a complex
output. We succeed the measurements of the real and imaginary
parts of the voltage in positive and negative frequencies.
Oscillations of the PT-dimers are also probed in the exact phase
of the Parity-Time symmetry without any restriction or
approximation. This work paves the way for the design of new
electronic and optolectronic devices. It opens the investigation
on the exploitation of the new band of negative frequencies in
Telecommunications and the generation of new oscillators for
simulations of qubits in quantum computing.


\begin{thebibliography}{}
\bibitem{a1} C. M. Bender and S. Boettcher, Real Spectra in Non-Hermitian Hamiltonians Having PT-Symmetry, Physical Review Letters, vol. 80, no. 24, pp. 5243-5246, Jun. 1998.
\bibitem{a2} C. M. Bender, S. Boettcher, and P. N. Meisinger, PT-symmetric quantum mechanics,  Journal of Mathematical Physics, vol. 40, no. 5, pp. 2201-2229, May 1999.
\bibitem{a3} C. M. Bender, M. V.Berry, and A. Mandilara, Generalized PT symmetry and realspectra, Journal of Physics A: Mathematical and General, vol. 35,no. 31, pp. L467-L471, Jul. 2002.
\bibitem{a4} A. Mostafazadeh, Pseudo-Hermiticity versus PT symmetry: The necessary condition for the reality of the spectrum of a non-Hermitian Hamiltonian, Journal of Mathematical Physics, vol.43, no. 1, pp. 205-214, Jan. 2002
\bibitem{a5} A. Mostafazadeh and A. Batal, Physical aspects of pseudo-Hermitian and PT-symmetric quantum mechanics, Journal of Physics A: Mathematical and General, vol. 37, no. 48, pp. 11645–11679, Nov. 2004.
\bibitem{a6} A. Mostafazadeh, Application of pseudo-Hermitian quantum mechanics to a PT-symmetric Hamiltonian with a continuum of scattering states, Journal of Mathematical Physics, vol. 46, no. 10, p. 102108, Oct. 2005.
\bibitem{a7} C. E. Rüter, K. G. Makris, R. El-Ganainy, D. N. Christodoulides, M. Segev, and D. Kip, Observation of parity–time symmetry in optics, Nature Physics, vol. 6, no. 3, pp. 192–195, Jan. 2010
\bibitem{a8} S. Nixon and J. Yang, All-real spectra in optical systems with arbitrary gain-and-loss distributions, Physical Review A, vol. 93, no. 3, Mar. 2016.
\bibitem{a9} R. El-Ganainy, K. G. Makris, M. Khajavikhan, Z. H. Musslimani, S. Rotter, and D. N. Christodoulides, Non-Hermitian physics and PT symmetry, Nature Physics, vol. 14, no. 1, pp. 11-19, Jan. 2018.
\bibitem{a10} H. Zhao and L. Feng, Parity-time symmetric photonics, National Science Review, vol. 5, no. 2, pp. 183-199, Jan. 2018.
\bibitem{a11} M.-A. Miri and A. Alù, Exceptional points in optics and photonics, Science, vol. 363, no. 6422, p. eaar7709, Jan. 2019.
\bibitem{a12} K. Özdemir, S. Rotter, F. Nori, and L. Yang, Parity-time symmetry and exceptional points in photonics, Nature Materials, vol. 18, no. 8, pp. 783-798, Apr. 2019.
\bibitem{a13} K. G. Makris, R. El-Ganainy, D. N. Christodoulides, and Z. H. Musslimani, Beam Dynamics in PT-Symmetric Optical Lattices, Physical Review Letters, vol. 100, no. 10, Mar. 2008.
\bibitem{a14} P. Tassin, L. Zhang, T. Koschny, E. N. Economou, and C. M. Soukoulis, Low-Loss Metamaterials Based on Classical Electromagnetically Induced Transparency, Physical Review Letters, vol. 102, no. 5, Feb. 2009.
\bibitem{a15} A. Alù and N. Engheta, All Optical Metamaterial Circuit Board at the Nanoscale, Physical Review Letters, vol. 103, no.14, Sep. 2009.
\bibitem{a16} H. Alaeian and J. A. Dionne, Parity-time-symmetric plasmonic metamaterials, Physical Review A, vol. 89, no. 3, Mar. 2014.
\bibitem{a17} A. Baev, P. N. Prasad, H. Ågren, M. Samo, and M. Wegener, Metaphotonics: An emerging field with opportunities and challenges, Physics Reports, vol. 594, pp. 1-60, Sep. 2015.
\bibitem{a18} C. M. Bender, B. K. Berntson, D. Parker, and E. Samuel, Observation of PT phase transition in a simple mechanical system, American Journal of Physics, vol. 81, no. 3, pp. 173–179, Mar. 2013.
\bibitem{a19} E. N. Tsoy, Coupled oscillators with parity-time symmetry, Physics Letters A, vol. 381, no. 5, pp. 462–466, Feb. 2017.
\bibitem{a20} X.-W. Xu, Y. Liu, C.-P. Sun, and Y. Li, Mechanical PT-symmetry in coupled optomechanical systems, Physical Review A, vol. 92, no.1, Jul. 2015.
\bibitem{a21} W. Li, Y. Jiang, C. Li, and H. Song, Parity-time-symmetry enhanced optomechanically-induced-transparency, Scientific Reports, vol. 6, no. 1, Aug. 2016.
\bibitem{a22} L.-Y. He, Parity-time-symmetry–enhanced sideband generation in an optomechanical system, Physical Review A, vol. 99, no. 3, Mar. 2019.
\bibitem{a23} C. Jiang, Y. Cui, Z. Zhai, H. Yu, X. Li, and G. Chen, Tunable slow and fast light in parity-time-symmetric optomechanical systems with phonon pump, Optics Express, vol. 26, no. 22, p. 28834, Oct. 2018.
\bibitem{a24} X. Zhu, H. Ramezani, C. Shi, J. Zhu, and X. Zhang, PT-symmetric acoustics, Phys. Rev. X 4, 031042 (2014).
\bibitem{a25} R. Fleury, D. Sounas, and A. Alu, An invisible acoustic sensor based on parity-time symmetry, Nat. Commun. 6, 5905 (2015).
\bibitem{a26} C. Shi, M. Dubois, Y. Chen, L. Cheng, H. Ramezani, Y. Wang, and X. Zhang, Accessing the exceptional points of parity-time symmetric acoustics, Nat. Commun. 7, 11110 (2016).
\bibitem{a27} J. Schindler, A. Li, M. C. Zheng, F. M. Ellis, and T. Kottos, Experimental study of active LRC circuits with PT-symmetries, Physical Review A, vol. 84, no. 4, Oct. 2011.
\bibitem{a28} J. Schindler, Z. Lin, J. M. Lee, H. Ramezani, F. M. Ellis, and T. Kottos, PT-symmetric electronics, Journal of Physics A:Mathematical and Theoretical, vol. 45, no. 44, p. 444029, Oct. 2012.
\bibitem{a29} F. Fotsa-Ngaffo, S. B. Tabeu, S. Tagouegni, and A. Kenfack-Jiotsa, Thresholdless characterization in space and time reflection symmetry electronic dimers, Journal of the Optical Society of America B, vol. 34, no. 3, p. 658, Feb. 2017.
\bibitem{a30} S. B. Tabeu, F. Fotsa-Ngaffo, and A. Kenfack-Jiotsa, Non-Hermitian Hamiltonian of two-level systems in complex quaternionic space: An introduction in electronics, EPL(Europhysics Letters), vol. 125, no. 2, p. 24002, Feb. 2019.
\bibitem{a31} S. B. Tabeu, F.Fotsa-Ngaffo, and A. Kenfack-Jiotsa, Imaginary resistor based Parity-Time symmetry electronics dimers, Optical and Quantum Electronics, vol. 51, no. 10, p. 335, Oct. 2019.
\bibitem{a32} X. Wang and J.-H. Wu, Optical PT-symmetry and PT-antisymmetry in coherently driven atomic lattices, Optics Express, vol. 24, no.4, p. 4289, Feb. 2016.
\bibitem{a33} P. Peng, W. Cao, C. Shen, W. Qu, J.Wen, L. Jiang, and Y. Xiao, Antiparity–time symmetry with flying atoms, Nature Physics, vol. 12, no. 12, pp. 1139–1145, Aug. 2016
\bibitem{a34} C. Zheng, Duality quantum simulation of a generalized anti-PT-symmetric two-level system, EPL (Europhysics Letters), vol. 126, no. 3, p. 30005, Jun. 2019.
\bibitem{a35} B. Lu, X.-F. Liu, Y.-P. Gao, C. Cao, T.-J. Wang, and C. Wang, Berry phase in an anti-PT symmetric metal-semiconductor complex system, Optics Express, vol. 27, no. 16, p. 22237, Jul. 2019.
\bibitem{a36} Y. Li, Y. Peng, L. Han, M.-A. Miri, W. Li, M. Xiao, X.-F. Zhu, J. Zhao, A. Alù, S. Fan, C.-W. Qiu,  Anti–parity-time symmetry in diffusive systems Science   Vol. 364, Issue 6436, pp. 170-17312 Apr 2019:
\bibitem{a37} J. Zhao, Y. Liu, L. Wu, C.-K. Duan, Y. Liu, and J. Du, Observation of Anti- PT -Symmetry Phase Transition in the Magnon-Cavity-Magnon Coupled System,” Physical Review Applied, vol. 13, no. 1, Jan. 2020.
\bibitem{a38} J. Wen, G. Qin, C. Zheng, S. Wei, X. Kong, T. Xin, and G. Long, Observation of information flow in the anti-PT-symmetric system with nuclear spins, npj Quantum Information, vol. 6, no. 1, Mar.2020.
\bibitem{a39} K. Shouno, Y. Ishibashi, Synthesis and active realization of a three-phase complex coefcient flter using gyrators. In: 51st Midwest Symposium on Circuits and Systems IEEE MWSCAS 2008, vol.87, pp. 550–553 (2008)
\bibitem{r1}  S. B. Tabeu, F. Fotsa-Ngaffo, and A. Kenfack-Jiotsa, Observation of Hermitian and Non-Hermitian Diabolic Points and Exceptional Rings in Parity-Time symmetric ZRC and RLC Dimers, arXiv:2005.11311 [physics.app-ph] (2020)
\bibitem{a40} P.-Y. Chen, M. Sakhdari, M. Hajizadegan, Q. Cui, M. M.-C. Cheng, R. ElGanainy, and A. Alù, Generalized parity–time symmetry condition for enhanced sensor telemetry,” Nature Electronics, vol.1, no. 5, pp. 297–304, May 2018.
\bibitem{a41} M. Hajizadegan, M. Sakhdari, S. Liao, and P.-Y. Chen, High-Sensitivity Wireless Displacement Sensing Enabled by PT-Symmetric Telemetry, IEEE Transactions on Antennas and Propagation, vol. 67, no. 5, pp. 3445– 3449, May 2019.
\bibitem{a42} L. O’Brien, Optical Quantum Computing, Science, vol. 318, no.5856, pp. 1567–1570, Dec. 2007.
\bibitem{a43} P. Kok, W. J. Munro, K. Nemoto, T. C. Ralph, J. P. Dowling, and G. J. Milburn, Linear optical quantum computing with photonic qubits, Reviews of Modern Physics, vol. 79, no. 1, pp. 135–174, Jan. 2007
\bibitem{a44} M. G. Thompson, A. Politi, J. C. F. Matthews, and J. L. O’Brien, Integrated waveguide circuits for optical quantum computing, IET Circuits, Devices  Systems, vol. 5, no. 2, p. 94, 2011
\bibitem{a45} C. Zheng, Duality quantum simulation of a general parity-time-symmetric two-level system, EPL (Europhysics Letters), vol. 123, no. 4, p. 40002, Sep. 2018


\end{thebibliography}
\end{document}